\documentclass[sigconf]{acmart}

\usepackage{tcolorbox}
\usepackage{multirow}
\usepackage{graphicx}
\usepackage[table,xcdraw]{xcolor}
\usepackage{makecell} 

\AtBeginDocument{%
  }

\setcopyright{acmlicensed}
\copyrightyear{2018}
\acmYear{2018}
\acmDOI{XXXXXXX.XXXXXXX}
\acmConference[Conference acronym 'XX]{Make sure to enter the correct
  conference title from your rights confirmation email}{June 03--05,
  2018}{Woodstock, NY}
\acmISBN{978-1-4503-XXXX-X/2018/06}

\begin{document}

\title{RefAgent: A Multi-agent LLM-based Framework for Automatic Software Refactoring}

\author{Khouloud Oueslati}
\email{khouloud.oueslati@polymtl.ca}
\orcid{0009-0002-9641-885X}
\affiliation{%
  \institution{Polytechnique Montreal}
  \city{Montreal}
  \country{Canada}
}

\author{Maxime Lamothe}
\email{maxime.lamothe@polymtl.ca}
\affiliation{%
  \institution{Polytechnique Montreal}
  \city{Montreal}
  \country{Canada}}

\author{Foutse Khomh}
\email{foutse.khomh@polymtl.ca}
\affiliation{%
  \institution{Polytechnique Montreal}
  \city{Montreal}
  \country{Canada}
}


\begin{abstract}
Recent advancements in Large Language Models (LLMs) have substantially influenced various software engineering tasks, including code generation, program repair, and software maintenance. Indeed, in the case of software refactoring, traditional LLMs have shown the ability to reduce development time and enhance code quality. However, these LLMs often rely on static, detailed instructions for specific tasks. In contrast, LLM-based agents can dynamically adapt to evolving contexts and autonomously make decisions by interacting with software tools and executing workflows. In this paper, we explore the potential of LLM-based agents in supporting refactoring activities. Specifically, we introduce RefAgent, a multi-agent LLM-based framework for end-to-end software refactoring. RefAgent consists of specialized agents responsible for planning, executing, testing, and iteratively refining refactorings using self-reflection and tool-calling capabilities.
We evaluate RefAgent on eight open-source Java projects, comparing its effectiveness against a single-agent approach, a search-based refactoring tool, and historical developer refactorings. Our assessment focuses on: (1) the impact of generated refactorings on software quality, (2) the ability to identify refactoring opportunities, and (3) the contribution of each LLM agent through an ablation study.
Our results show that RefAgent achieves a median unit test pass rate of 90\%, reduces code smells by a median of 52.5\%, and improves key quality attributes (e.g., reusability) by a median of 8.6\%. Additionally, it closely aligns with developer refactorings and the search-based tool in identifying refactoring opportunities, attaining a median F1-score of 79.15\% and 72.7\%, respectively. Compared to single-agent approaches, RefAgent improves the median unit test pass rate by 64.7\% and the median compilation success rate by 40.1\%. These findings highlight the promise of multi-agent architectures in advancing automated software refactoring.
\end{abstract}



\setcopyright{none} 
\settopmatter{printacmref=false} 
\renewcommand\footnotetextcopyrightpermission[1]{}
\maketitle
\section{Introduction}\label{sec:intro}
Large-scale software systems tend to increase in complexity and become difficult to maintain as they evolve to adapt to changing requirements \cite{LEHMAN198419}. Software refactoring is a standard practice that aims to enhance code design without altering its observable behavior \cite{Fowler1999}. 
Neglecting refactoring is often associated with the accumulation of technical debt, leading to an increase in code smells \cite{LACERDA2020110610, 7503738} or design flaws that violate design principles and compromise the code understandability and maintainability \cite{YAMASHITA20132639, 1173068}. These negative consequences ultimately lead to higher maintenance costs \cite{10.1147/sj.153.0225, Chen2016}. 

Since manual refactoring is error-prone, time-consuming, and ineffective when extensive changes are needed \cite{FERNANDES2020106347,inproceedings0x}, various semi-automated and fully-automated techniques have been proposed over the past decade \cite{10418470, 6032586, MORALES2017236}. While IDEs offer built-in refactoring options, they support only a limited set of refactorings and still require manual effort. A Microsoft survey found that 28\% of developers face challenges with refactoring, especially in large codebases and ensuring correctness \cite{6802406}. Fully-automated techniques, such as search-based refactoring, offer an appealing alternative by formulating refactoring as an optimization problem to identify a refactoring sequence that enhances the program based on a defined fitness function, which can involve factors such as code smells or software quality metrics~\cite{10.1016/j.jss.2015.03.040}, but they often radically change program design \cite{articlevahid}, lack behavior preservation guarantees\cite{10.1007/s11219-019-09477-y,alomar2021preservingbehaviorsoftwarerefactoring}, and support a limited range of refactoring types, and are computationally prohibitive for large projects \cite{10.1145/2541348.2541357, Ecole2878}. 

Recent breakthroughs in Large Language Models (LLMs) have sparked growing interest in applying generative AI to software engineering tasks such as code generation \cite{jiang2024surveylargelanguagemodels}, fault localization \cite{10.1145/3638530.3664174}, and automatic program repair \cite{li2024hybridautomatedprogramrepair}. The emergence of LLM-based commercial tools like GitHub Copilot\footnote{https://github.com/features/copilot} and AutoCoderRover\footnote{https://autocoderover.dev/} highlights practitioners' growing interest in LLM-powered automated software development. Developer adoption of AI is rising \cite{daigle2024aiwave}, with 76\% using or planning to use it, and 92\% having tried tools like Copilot. Early efforts have also explored the use of LLMs in refactoring \cite{cordeiro2024empiricalstudycoderefactoring, liu2024empiricalstudypotentialllms}, showing promise in reducing developer effort and improving design quality. However, despite this progress, fully automated refactoring remains a largely unsolved problem. Existing techniques often suffer from one or more of the following limitations: poor scalability to large codebases, insufficient coverage of refactoring types, difficulty in maintaining behavioral correctness, and brittleness when applying transformations in real-world systems \cite{6802406, 10.1145/2541348.2541357, Ecole2878}.

This persistent gap highlights the need for more dynamic, context-aware, and adaptive solutions that can not only propose refactorings but also coordinate complex, multi-step workflows involving validation, correction, and iterative refinement. In this context, LLM-based agents, defined as entities that use LLMs as the cognitive core of the agent,  can dynamically adapt to changing contexts, interact with tools, and execute workflows to achieve goals \cite{he2024llmbasedmultiagentsystemssoftware}, offer a promising new direction for this challenge, not just because LLMs are powerful, but because agents can be structured to simulate the reasoning and actions of skilled developers, distributing responsibilities such as code analysis, transformation, compilation, and testing across specialized roles \cite{liu2024largelanguagemodelbasedagents}. Compared to static prompt-based LLMs, agents can reason, respond to intermediate feedback (e.g., test failures or compilation errors), and adjust plans dynamically. They can generate code, reflect on its consequences, revise decisions, and collaborate with one another, mirroring the way human developers iteratively and interactively perform refactoring in practice. Furthermore, agents enable modular reasoning: by breaking the refactoring process into smaller, goal-directed tasks, which reduces cognitive and computational complexity, improving both accuracy and interpretability \cite{wang2024agentssoftwareengineeringsurvey, xia2024agentlessdemystifyingllmbasedsoftware}.

Critically, multi-agent systems bring an additional advantage: coordination across specialized roles. A single LLM may hallucinate over prolonged interactions, but a system of agents, each focused on a well-defined subtask, can collaboratively handle complexity through role specialization and communication. This design aligns closely with how real-world refactoring is performed: as a sequence of discrete, dependent steps that require both autonomy and collaboration \cite{he2024llmbasedmultiagentsystemssoftware}.

In this paper, we introduce RefAgent, a novel multi-agent, LLM-based, fully automated framework for software refactoring. Our approach simulates the sequential nature of software refactoring workflows by distributing tasks among agents, reducing development time, mitigating hallucinations, and ensuring behavior preservation in complex software engineering environments. Our goal in creating and evaluating this approach is to determine the current status of the potential of LLM-based multi-agents for software refactoring. RefAgent consists of four key components:


\textit{Context-Aware Planner Agent} – Identifies refactoring opportunities and generates a structured plan based on dependency analysis and code metrics.

\textit{Refactoring Generator Agent} – Executes the refactoring plan provided by the Context-Aware Planner Agent on the target class, producing a refactored version of that class.

\textit{Compiler Agent} – Interacts with the compilation environment and the Refactoring Generator Agent through an iterative feedback loop to ensure any compilation issues are addressed.

\textit{Tester Agent} – Ensures that the refactored target class preserves functionality using existing and auto-generated tests using Evosuite, cooperating with the Refactoring Generator Agent to fix test failures.





To evaluate the effectiveness of RefAgent, we ask the following research questions:

\textit{\textbf{RQ1: How effective is our approach in improving the quality of software projects?}} 
To evaluate the effectiveness of RefAgent in enhancing the quality of software projects, we measure code Smells Reduction Rates (SRR) along with compilation success rates, and unit test pass rates across eight software projects. 
Furthermore, we examine the prevalent refactoring types and discuss their implications for future improvements.

\textbf{\textit{RQ2: How effective is our approach in identifying refactoring opportunities and improving software quality compared to search-based techniques and developers?}} We assess the ability of RefAgent in identifying refactoring opportunities by comparing its refactoring patches across code regions against RefGen, a search-based refactoring tool, and developers using precision, recall, and F1-score. Moreover, we compare their software quality improvement rates using the Quality Model for Object-Oriented Design (QMOOD) metrics across eight software projects. 

\textbf{\textit{RQ3: What is the contribution of each component of our framework? (Ablation study)}} To further showcase the effectiveness of RefAgent in improving software quality, we reuse the performance criteria from RQ1 to compare the performance of RefAgent against single LLM-based approaches. Moreover, we investigate the contribution of key components in RefAgent by conducting an ablation study to evaluate the performance under different settings, focusing on the \textit{Context-aware Planner Agent} and the impact of the iterations in the feedback loops.

Overall, our findings underscore the potential of multi-agent LLM architectures for advancing automated refactoring. 

\section{Background and Related Work}\label{sec:background}

Recent advancements in Large Language Models (LLMs) have substantially influenced various software engineering tasks \cite{yang2024chainofthoughtneuralcodegeneration,tian2024fixinglargelanguagemodels}. Particularly, in the case of software refactoring, an empirical study by Cordeiro et al. \cite{cordeiro2024empiricalstudycoderefactoring} evaluated the refactorings generated by StarCoder2 \cite{lozhkov2024starcoder2stackv2} against developers. They apply zero-shot and one-shot prompting and assess the unit test pass rate using the pass@k metric. Their results show that LLMs effectively reduce code smells and apply various refactoring types. While their study focused on a single LLM with a static prompt at commit-level granularity, we introduce a fully automated multi-LLM agent approach that leverages specialized LLM agents with tool-calling capabilities to retrieve context, interact with compilation and testing environments, and perform complex, context-aware refactorings without manual intervention.

Furthermore, we base our approach on the work proposed by Choi et al. \cite{10.1007/978-3-031-64573-0_4}. The authors present a single LLM approach that iteratively refactors methods identified as having high cyclomatic complexity by refining the LLM's suggestions using the error stack trace from compilation and testing environment across 20 iterations. However, their pipeline is limited due to sequential refactoring of individual methods, while our approach refactors the projects by iterating through all classes of the project. In addition, we decentralize the process by assigning distinct roles to specialized LLM agents, each responsible for a specific subtask and environment (e.g., compilation or testing). This modular design ensures that each agent receives and processes feedback (e.g., stack traces) within its relevant execution context. Moreover, the proposed pipeline by Choi et al. lacks contextual input for the LLM to generate the refactoring. Yet, refactoring is a context-dependent problem \cite{MORALES2017236}, thus we employ a context retrieval agent to help the LLM decide and plan the refactorings effectively. 



Multi-agent systems have shown promise in software maintenance by leveraging specialized agents that work collaboratively or competitively to achieve a final goal while invoking external tools to receive context and feedback to solve issues in real-world software projects. For example, agents like AutoCodeRover \cite{zhang2024autocoderoverautonomousprogramimprovement} and Masai \cite{arora2024masaimodulararchitecturesoftwareengineering} use static and dynamic checks to validate generated patches, while others invoke tools such as those for syntactic correctness checking, code format checking \cite{ma2025alibabalingmaagentimprovingautomated}, and vulnerability detection \cite{nunez}. However, the use of multi-agent systems for end-to-end refactoring remains largely unexplored.

\begin{figure*}
    \centering
    \includegraphics[width=\linewidth]{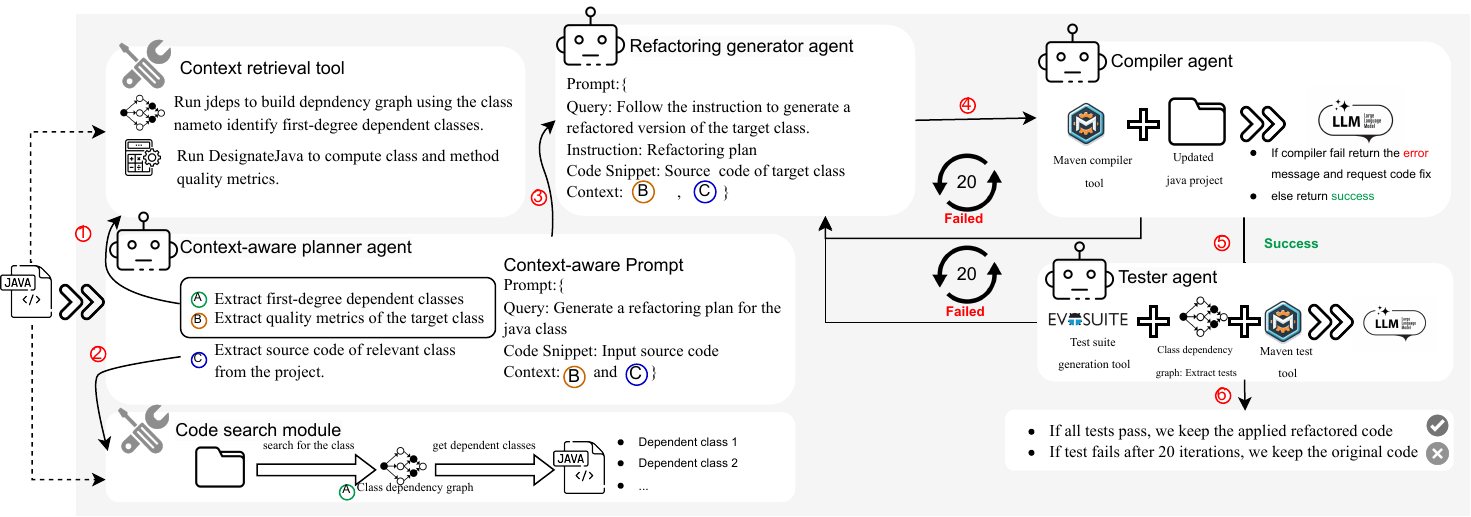}
    \vspace{-8mm}
    \caption{RefAgent Overview}
    \vspace{-4mm}
    \label{fig:refagent}
\end{figure*}

\section{RefAgent Approach}\label{sec:approach}
\subsection{Overview}
In this section, we present an overview of the overall phases of RefAgent, a multi-agent framework that iteratively improves the quality of each class in a given Java project. RefAgent is designed to simulate the sequential nature of end-to-end refactoring, where the goal is to improve weakly structured code through extensive refactoring, commonly known as Root-canal refactoring \cite{FERNANDES2020106347}. The framework allows users to run the workflow autonomously to enhance the quality of a given codebase, significantly reducing the manual effort required for large-scale refactoring tasks. For a given project, a randomly-selected target class as a starting point, the workflow of RefAgent leverages specialized LLM agents for the context-aware identification of refactoring opportunities and planning as well as ensuring the compilation of the modified source code, and thorough testing to ensure that functionality is preserved for the modified target class. 

Particularly, following previous work \cite{10.1007/978-3-031-64573-0_4}, RefAgent iteratively analyzes feedback across 20 iterations using external tools such as the compilation and testing environments, and integrates linguistic reflections to refine its refactoring suggestions \cite{shinn2023reflexionlanguageagentsverbal}. This iterative self-improvement mechanism enables RefAgent to dynamically adapt and enhance the quality of code transformations. The following sections provide detailed descriptions of each of our four agents.

\subsection{Context-aware Planner Agent}

RefAgent employs an LLM agent, the \textbf{Context-aware Planner Agent}, which uses a Java class as a starting point. The agent is equipped with tool-calling capabilities that allow the model to detect when one or more tools should be called and respond with the inputs that should be passed to those tools. Specifically, it can call the context-retrieval module that runs the dependency analysis tool \textit{jdeps}, passing the class name and the project path as inputs, which generates a class-level dependency graph. Next, following prior work \cite{cordeiro2024empiricalstudycoderefactoring}, the agent invokes a code smell extraction tool \textit{DesigniteJava 2.5.2}~\cite{Sharma2024} that also provides software metrics detailed in Section~\ref{sec:code_metrics}. DesigniteJava then takes the path to the source Java code as input and returns a set of quality metrics (e.g, cyclomatic complexity, lack of cohesion, etc.) that the Context-aware Planner Agent will use for identifying refactoring opportunities and generating a refactoring plan for the target class. 
Note that within RefAgent, we do not use its code smell detection capabilities in the refactoring process itself. Instead, the planner agent leverages only the extracted code metrics to provide context for the decision-making driven by the agent's reasoning. We ensure that the planner's decisions about which code regions to refactor are not biased by labeled smell instances from DesigniteJava.
\begin{figure}[H]
    \vspace{-4mm}
    \centering    \includegraphics[width=\linewidth]{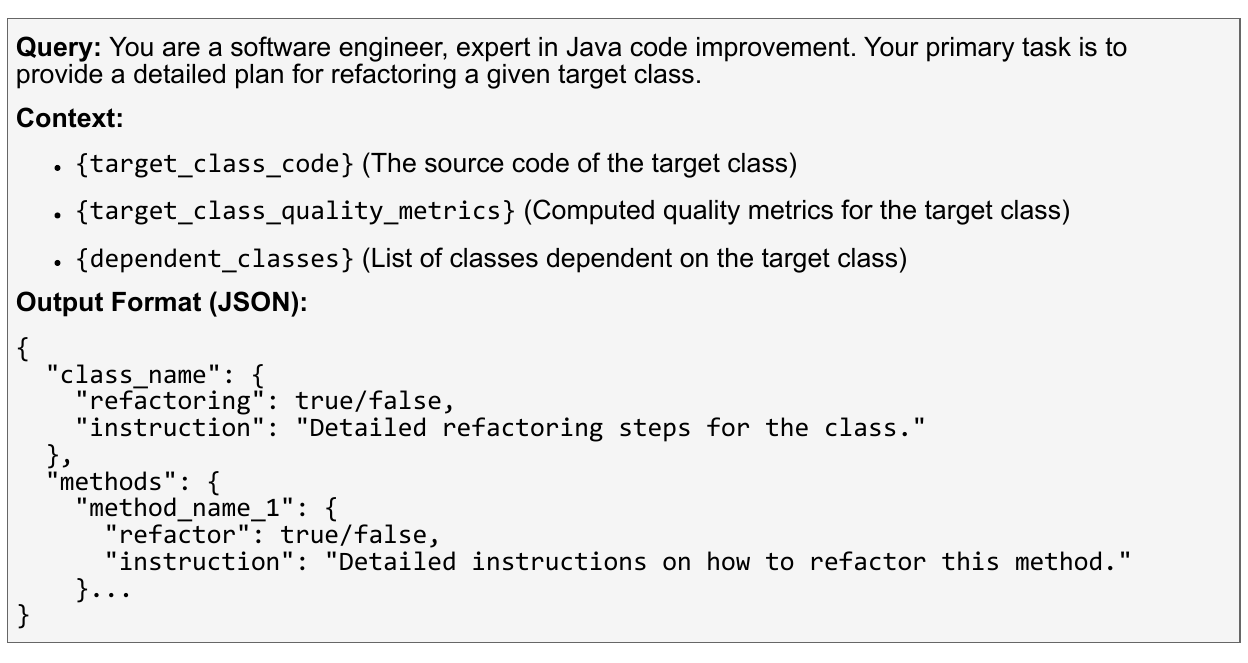}
    \vspace{-7mm}
    \caption{Context-aware Planner Prompt}
    \vspace{-4mm}
    \label{fig:planner}
\end{figure}


After, the Context-aware Planner Agent calls the \textit{code-search module}, which is a module that retrieves the source code of the target class as well as the source code of its first-degree dependent classes. The module searches for the target source code and its direct dependencies by taking as input the full name of the target class (including its package) and the dependency graph provided by the jdeps tool. It then iterates through the project folder to locate and return the source code of the required classes. Therefore, the agent considers information from related classes to suggest higher-level design improvements or class-level refactorings when applicable.

Next, the Context-aware Planner Agent builds the prompt according to the template as shown in Figure~\ref{fig:planner} for its core LLM to identify refactoring opportunities and assign the proper refactorings for the various code regions (e.g., method, field, variable) of the target class. 
Finally, the Context-aware Planner Agent provides a refactoring plan comprising of the particular code regions identified for improvement along with explicit refactoring instructions designed to be interpretable by downstream LLM agents.

\subsection{Refactoring Generator Agent}

The primary objective of the \textbf{Refactoring Generator Agent} is to generate refactored Java code following the instructions of  Context-aware Planner Agent, while ensuring the correctness and functionality of the output.  

This agent takes as input the original source code of the target class, the source code of dependent classes, software metrics, and the refactoring plan. It applies the refactoring plan to produce an improved version of the class, updates the code, and initiates the compilation phase. The Refactoring Generator Agent should return Java code as output as instructed in the prompt. 

As shown in Figure \ref{fig:refagent}, the Refactoring Generator Agent is invoked by different agents, including the Compiler Agent and the Tester Agent, in case of compilation or test failures in order to dynamically adapts its prompt while maintaining the primary objective of following the refactoring plan as shown in the Figure \ref{fig:planner}. The prompt context can vary depending on:

– Compilation errors (when called by the Compiler Agent)

– Test failures (when called by the Tester Agent)

– The original refactoring plan (when initially generating the refactored class)  


Regardless of the context, the Refactoring Generator Agent is required in the prompt to ensure that the refactoring instructions from the planner remain the guiding principle. It continuously refines the code while ensuring that the final output is syntactically correct, functional, and successfully passes compilation and testing.

\begin{figure}[H]
    \vspace{-2mm}
    \centering    \includegraphics[width=\linewidth]{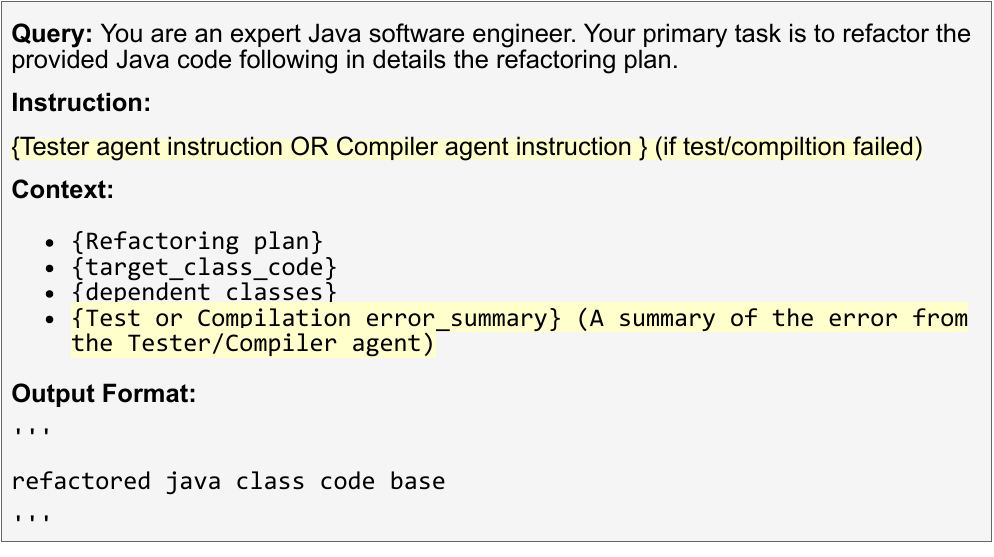}
    \vspace{-5mm}
    \caption{Illustrative example of the Refactoring Generator Agent prompt}
    \vspace{-4mm}
    \label{fig:generator}
\end{figure}
\subsection{Compiler Agent}
The \textit{Compiler Agent} employs a self-reflection loop \cite{shinn2023reflexionlanguageagentsverbal} by interacting with the Maven compiler tool to validate the syntax of the refactored code. If a compilation error occurs, the error message is sent to the LLM, which analyzes the error, generates an error summary, and forwards it to the Refactoring Generator Agent for reflection and correction. The feedback loop is set to a maximum of 20 iterations, following previous work \cite{10.1007/978-3-031-64573-0_4} to assess the extent to which it can effectively refine the generated refactorings. The Compiler Agent’s reflection process is as follows:

\begin{enumerate}
    \item After the Refactoring Generator agent proposes a refactoring, the compiler agent invokes the Maven compiler tool to compile the updated refactored project. This tool executes the Maven compilation command on the operating system without running tests, ensuring that the code syntax is correct.
    \item The Maven compiler tool then returns logs to the compiler agent, detailing the compilation status.

    – If compilation succeeds (100\%), the compiler agent confirms success and proceeds to the Tester agent for further validation.

    – If compilation fails, the compiler agent analyzes the error logs, extracts the necessary debugging information, and generates a structured error summary.

    \item The error summary, along with specific fixing instructions, are sent to the Refactoring Generator Agent, enabling it to reflect on the error, modify the code, and attempt a fix. This iterative process continues until successful compilation is achieved or the iteration limit is reached.
\end{enumerate}
\vspace{-3mm}
\subsection{Tester Agent}
Code refactoring should not alter the external functionality of the code \cite{Fowler1999}. It is thus necessary to execute tests to evaluate whether the refactored code retains its original functionality. In this process, the \textit{Tester Agent} uses the code search module to locate developer-written tests from the class-level dependency graph generated by \textit{jdeps} and employs the automated test suite generation tool \textit{EvoSuite} to generate additional regression tests, The regression tests are created based on the original program before any refactoring is applied. We use these tests in an attempt to maximize the validation of the modified source code.

Similar to the Compiler Agent, the Tester Agent employs a feedback loop of up to 20 iterations. If a test fails, it reads the logs, analyzes the errors, and generates a summary report. This report, along with the failing test cases, is sent to the Refactoring Generator agent, requesting a fix as shown in Figure~\ref{fig:generator}. If the refactoring patch continues to fail after all iterations, the agent excludes the target class from further improvements and proceeds to the next iteration without applying the refactorings.
The Tester agent interacts with three main tools:

\textbf{\textit{EvoSuite.}}
Since some projects may lack developer-written test cases, Tester Agent calls \textit{EvoSuite} which takes the target class name as input and automatically generates unit tests optimized for maximum code coverage (e.g, Line Coverage, Branch Coverage and Output Coverage). EvoSuite integrates \textit{JUnit 4} assertions to capture and validate expected behavior in test cases. 



\textbf{\textit{JDeps.}}
The \textit{jdeps} tool extracts the dependency graph of the project and identifies test cases that are directly related to the target classes. This ensures that relevant developer-written tests are identified and executed for validation.

\textbf{\textit{Maven Test Tool.}} Similar to the Maven compiler tool, the Maven test tool interacts with the operating system to execute tests using Maven. The Tester Agent triggers the execution of both developer-written and EvoSuite-generated tests to ensure that the observable behavior of the refactored code remains unchanged. 

Finally, this workflow is autonomously executed throughout all the classes of the project.
In Section \ref{sec:results}, we evaluate the impact of RefAgent on 8 software projects using various evaluation metrics. We compare RefAgent against developers, a search-based refactoring tool named RefGen \cite{MORALES201825}, and single-agent scenarios. We also examine the performance of our framework RefAgent, using three LLMs, notably GPT-4o, StarCoder2 and DeepseekCoder.


\section{Experimental Methodology}\label{sec:methodology}
\subsection{Studied Models and Dataset}

\textbf{Models.} We initially design RefAgent using the closed-source GPT-4o model as its core LLM. To further assess its performance, we benchmark RefAgent against two open-source models: StarCoder2-15B-instruct \cite{lozhkov2024starcoder2stackv2} and DeepSeekCoder-33B-instruct \cite{deepseekcoder2024}. Prior work highlights StarCoder2’s effectiveness in refactoring, showing improvements in code quality and unit test pass rates using zero-shot and one-shot prompting \cite{cordeiro2024empiricalstudycoderefactoring}. The study also mitigates data leakage by filtering recent Java projects not present in the StackV2 training data \cite{10.1145/3379597.3387487, lozhkov2024starcoder2stackv2}, an approach we follow in our setup.

DeepSeekCoder has demonstrated strong performance on code generation and reasoning benchmarks. We include it to evaluate whether large instruction-tuned open models can match or outperform GPT-4o in automated refactoring.

GPT-4o supports a 128K token context window, while StarCoder2 and DeepSeekCoder are limited to 8192 tokens. To ensure fair comparison, we run RefAgent only on classes under 4096 tokens, reserving the remaining space for dependent code and agent instructions, similar to OpenAI’s dynamic context handling in ChatGPT. We use a temperature of 0.7 for GPT-4o to balance consistency and diversity.

\begin{table}[H]
\centering
\vspace{-4mm}
\caption{Dataset Overview}
\label{tab:projects}
\vspace{-2mm}
\scalebox{0.73}{
        \begin{tabular}{|l|l|l|c|c|c|}
        \hline
        \rowcolor[HTML]{EFEFEF} 
        Project name &
          Release &
          Release date &
          \multicolumn{1}{l|}{\cellcolor[HTML]{EFEFEF}Nbr of classes} &
          \multicolumn{1}{l|}{\cellcolor[HTML]{EFEFEF}Nbr of Methods} &
          \multicolumn{1}{l|}{\cellcolor[HTML]{EFEFEF}KLOC} \\ \hline
        JClouds      & 2.3.0     & 05-2024 & 9,972 & 44,797 &  646 \\ \hline
        Accumulo     & 1.10.4    & 11-2023 & 2,286 & 40,353 & 603                      \\ \hline
        systemml     & 3.2.0-rc1 & 02-2024 & 3,986 & 41,962 & 658                      \\ \hline
        apex-core    & 3.7.0-rc1     & 07-2021 & 1,378 & 5,432  & 107                      \\ \hline
        skywalking   & 9.7.0     & 11-2023 & 2,871 & 9,736  & 192                      \\ \hline
        deltaspike   & 1.9.6     & 04-2022 & 2,155 & 6,471   & 139                      \\ \hline
        Jmeter       & 5.6.3-rc1 & 12-2023 & 1,723 & 14,074 & 247                      \\ \hline
        openmeetings & 7.2.0 & 12-2023 & 1,306 & 8,441  & 188                      \\ \hline
        \end{tabular}
}
\vspace{-2mm}
\end{table}

\textbf{Dataset.} We randomly select eight open-source Apache Java projects, due to resource constraints, from a dataset widely used in prior research \cite{10.1145/3379597.3387487, cordeiro2024empiricalstudycoderefactoring}. We then filter out projects that are included in StarCoder2's training dataset Stackv2 \cite{lozhkov2024starcoder2stackv2} to remove the risk of data-leakage on that model. Unfortunately, due to the closed-source nature of GPT-4o, we cannot ensure no data leakage for that model. We therefore rely on StarCoder2 to determine the strength of our approach in a more controlled setting. The details of the selected projects, including their size and characteristics, are summarized in Table~\ref{tab:projects}. 

\textbf{Hardware and Computational Resources.} StarCoder2 experiments are conducted on a computing server with 80 GB of memory. The operating system used is Ubuntu 22.04.4 LTS. We use a GPU-accelerated setup similar to previous work \cite{cordeiro2024empiricalstudycoderefactoring}, leveraging the Nvidia A100 GPUs for efficient model inference and code generation. The GPT4o-based experiments use GPT-4o via the OpenAI API, using cloud-based inference for model execution.

\subsection{Impact on Software Quality Assessment}
\subsubsection{\textbf{Code Smells Extraction}}
\label{sec:code_smells}

Code smells are indicators of underlying design or implementation issues that may impact maintainability, readability, and overall software quality \cite{10.5555/2755629}. Since code refactoring is commonly associated with the elimination of code smells \cite{YAMASHITA20132639}, we extract code smells from the Java projects to measure the capabilities of RefAgent in reducing code smells. We use \textit{DesigniteJava 2.5.2} \cite{Sharma2024} to extract code smells from the source code before and after the project is refactored. DesigniteJava detects 46 different types of code smells. We assess how well RefAgent mitigates code smells compared to competing approaches. we focus on the following categories:


\noindent
\textbf{Design Smells}: Poor adherence to design principles that hinder modularity and reusability (e.g., \textit{unnecessary abstraction}).

\noindent
\textbf{Implementation Smells}: Code-level issues that make the code harder to maintain (e.g., \textit{large classes}, \textit{long methods}).


\subsubsection{\textbf{Quality Metrics Computation}}
\label{sec:code_metrics}

\begin{table*}[h]
\centering
\caption{QMOOD Computation Equations.}
\vspace{-3mm}
\begin{tabular}{ll}
\hline
\rowcolor[HTML]{EFEFEF} 
\textbf{Quality Attribute} & \textbf{Quality Attribute Calculation}                     \\ \hline
Reusability                & -0.25 * DCC + 0.25 * CAM + 0.5 * CIS + 0.5 * DSC           \\ \hline
Flexibility                & 0.25 * DAM - 0.25 * DCC + 0.5 * MOA + 0.5 * NOP            \\ \hline
Understandability &
  \begin{tabular}[c]{@{}l@{}}-0.33 * ANA + 0.33 * DAM - 0.33 * DCC + 0.33 * CAM - 0.33 * NOP - 0.33 * NOM - 0.33 * DSC + 0.33 * CAM \\ - 0.33 * NOP - 0.33 * NOM - 0.33 * DSC\end{tabular} \\ \hline
Effectiveness              & 0.2 * ANA + 0.2 * DAM + 0.2  * MOA + 0.2 * MFA + 0.2 * NOP \\ \hline
Extendibility              & 0.5 * ANA - 0.5 * DCC + 0.5 * MFA + 0.5 * NOS              \\ \hline
Functionality              & 0.12 * MOA + 0.22 * MOP + 0.22 * CIS + 0.22 * DSC + 0.22 * NOH \\ \hline
\multicolumn{2}{l}{\begin{tabular}[c]{@{}l@{}}\textbf{Note:} DSC is design size, NOM is number of methods, DCC is coupling, NOP is polymorphism, NOH is number of hierarchies, \\ CAM is cohesion among methods,  ANA is avg. num. of ancestors, DAM is data access metric, MOA is measure of aggregation, \\ MFA is measure of functional abstraction,  and CIS is class interface size.\end{tabular}}
\end{tabular}
\label{tab:qmood}
\vspace{-4mm}
\end{table*}

We use the \textit{Quality Model for Object-Oriented Design (QMOOD)}, proposed by the prior work \cite{979986}, which consists of a set of quality measures using the ISO 9126 specification. It defines six high-level design quality attributes: reusability, flexibility, understandability, functionality, extensibility, and effectiveness that can be calculated using 11 lower-level design metrics as detailed in Table~\ref{tab:qmood}. We use QMOOD to estimate the effect of the suggested refactoring solutions on quality attributes, similarly to many prior works \cite{articlevahid, Ecole2878}. Likewise, we calculate the Quality Improvement (QI) for each quality attribute using the following formula.

\begin{equation}
QI(A_q) = \frac{A_q(p') - A_q(p)}{|A_q(p)|} \times 100
\end{equation}

\noindent where \( A_q(p) \) represents the measurement of quality attribute \( q \) for project \( p \), and \( p' \) denotes the refactored version of project \( p \).
The sign indicates an increase (+) or decrease (-), while the numerical value represents the percentage of improvement. Since QMOOD attribute calculations may yield negative values in the original design, taking the absolute value of the divisor is essential. 



By analyzing each QMOOD attribute before and after refactoring, we assess the effectiveness of RefAgent in enhancing software maintainability, modularity, and overall design quality. We compute the metrics using the tool available online.\footnote{\url{https://github.com/dimizisis/metrics_calculator/}}

\subsection{Identification of Refactoring opportunities}
\label{sec:refactoring}

\subsubsection{\textbf{Experimental details}}

We employ RefactoringMiner 3.0 \cite{9136878, 8453111}, which is a tool designed only to detect and classify refactoring operations between version histories or commits, to examine the refactorings applied by RefAgent. RefactoringMiner represents
the state of the art in refactoring detection, supporting up to 59 different types of refactorings.  The evaluation of  RefactoringMiner on this dataset, performed in previous
work \cite{10.1109/TSE.2023.3326775}, shows an overall precision of 99.7\% and a recall of 94.2\%, confirming it as an accurate refactoring detection tool \cite{10.1145/3180155.3180206}.
The process is as follows: for each project \( p \) at release version \( v \), we create a fork. After RefAgent refactors a class, we commit the modified class to the fork. 

For each commit corresponding to a refactored class in \( p \), we apply RefactoringMiner, which extracts the refactoring types and their locations within the commit. RefactoringMiner returns detailed information, including the class name, method name, and the specific lines of code where each refactoring was applied. We leverage this information to match RefAgent’s refactorings with the extracted refactorings, allowing us to compare our results with developer-introduced changes. Furthermore, for each project \( p \), we clone the next release \( v_{\text{n+1}} \) and apply RefactoringMiner. RefactoringMiner automatically extracts all commits in version \( v_{\text{n+1}} \) and returns the identified refactoring types and their locations. If no refactoring is found, it generates an empty folder containing only the commit ID. Note that in our study, RefactoringMiner is used exclusively in the evaluation of RefAgent. 
\subsubsection{\textbf{Relevance to Search-based Refactoring Approaches and Developers}}

Previous work has shown that search-based refactoring approaches excel in identifying refactoring opportunities by using optimization algorithms that explore the search space of possible refactorings to maximize software quality \cite{ALDALLAL2015231}. Thus, we compare the refactoring opportunities identified by RefAgent with those made by \textit{RefGen}, a search-based refactoring tool that implements efficient refactoring scheduling based on partial order reduction \cite{MORALES201825}. RefGen was initially developed as an Eclipse plug-in designed to suggest sequences of refactorings to improve the design quality of software systems by addressing anti-patterns and code smells detected across different classes. The generated refactoring sequence is ordered to maximize design quality improvement while avoiding conflicts among the suggested refactorings.

RefGen provides an option to run the tool in simulation mode. In this mode, RefGen constructs an abstract code design model, identifies anti-patterns, and generates a sequence of refactoring candidates, which are then applied to the design model rather than the actual code. This simulation capability is particularly useful for our analysis, as it allows us to evaluate the proposed refactoring solutions and estimate the potential design quality improvements resulting from applying a complete refactoring sequence.

The objective is to determine whether RefAgent applies refactorings in the same class, method, and location as baseline approaches, notably the developers and the RefGen tool.

To quantify the alignment between RefAgent and baseline approaches, we compute the following evaluation metrics:

\noindent \textit{Precision}: Measures the proportion of refactorings suggested by RefAgent that are also present in the selections done by baseline approaches.
\begin{equation}
Precision = \frac{TP}{TP + FP}
\end{equation}

\noindent \textit{Recall}: Measures the proportion of the baseline's refactorings that were also applied by RefAgent.
\begin{equation}
Recall = \frac{TP}{TP + FN}
\end{equation}

\noindent \textit{F1-score}: The harmonic mean of precision and recall, balancing correctness and completeness.
\begin{equation}
F1 = 2 \times \frac{Precision \times Recall}{Precision + Recall}
\end{equation}
A higher F1-score indicates strong alignment between RefAgent and developers, or between RefAgent and RefGen tool,  in identifying refactoring opportunities.

\section{Results}\label{sec:results}
\subsection{RQ1: How effective is our approach in improving the quality of software projects}
\label{RQ1}


\subsubsection{\textbf{Motivation}}
Improving software quality is a core objective of automated refactoring, yet existing techniques exhibit limitations such as producing behavior-breaking changes and a limited set of supported refactoring types. To evaluate whether RefAgent can address these limitations, we examine its impact on key quality indicators across 8  real-world software projects and discuss its prevalent refactoring types.

\subsubsection{\textbf{Approach}}

As indicated in  Section \ref{sec:methodology},  
we evaluate RefAgent using GPT-4o, DeepSeek-Coder and StarCoder2. Next, RefAgent runs autonomously with LLM agents collaborating towards a shared goal, which is refactoring the codebase by iterating through classes while preserving the behavior. As detailed in Section~\ref{sec:approach}, RefAgent starts by identifying refactoring opportunities, planning the appropriate refactoring solutions, and executing the rest of the workflow.

To evaluate the effectiveness of \textbf{RefAgent} in improving software quality, we measure its impact on \textbf{code smells, compilation success rates, and unit test pass rates}. First, a Java project is selected, then we use \textit{DesigniteJava} to detect and record the code smells. For the purpose of our study, we manually ensure that the project before refactoring is initially compilable and has a 100\% unit test pass rate with regards to its existing developer-written test suites.

The gain in each metric is assessed using the Improvement Rate (IR), following the approach used in previous studies~\cite{cordeiro2024empiricalstudycoderefactoring}. It is calculated as follows:

\begin{equation}
IR = \frac{m_{\text{before}} - m_{\text{after}}}{m_{\text{before}}} \times 100
\end{equation}

\noindent where \( m_{\text{before}} \) represents the initial metric value (e.g., code smell count or quality metric), and \( m_{\text{after}} \) represents the corresponding metric value after refactoring. This formula allows us to quantify the relative improvement introduced by RefAgent across different software quality dimensions.

Furthermore, we identify code smells in both the original and refactored projects, as described in Section~\ref{sec:code_smells}. We compute the Improvement Rate (IR) for each code smell type, grouping them by category to analyze which type is most effectively reduced by RefAgent. 

To assess the statistical differences in improvements for code smells and unit tests relative to the baseline across all projects, we use the Wilcoxon signed-rank test \cite{10.1145/1985793.1985795}, as the measurements being compared are paired. A significance threshold of p $< 0.05$ is used to determine statistical significance.

.
\begin{table}[H]
\centering
\vspace{-4mm}
\caption{RefAgent Performance Evaluation Across 8 Projects}
\vspace{-2mm}
\label{tab:refactoring_comparison}
\scalebox{0.82}{
    \begin{tabular}{|c|cc|cc|cc|}
        \hline
        \rowcolor[HTML]{EFEFEF} 
        \cellcolor[HTML]{EFEFEF} &
          \multicolumn{2}{c|}{\cellcolor[HTML]{EFEFEF}\textbf{\begin{tabular}[c]{@{}c@{}}Unit Test\\ Pass Rate\end{tabular}}} &
          \multicolumn{2}{c|}{\cellcolor[HTML]{EFEFEF}\textbf{\begin{tabular}[c]{@{}c@{}}Compilation \\ Pass Rate\end{tabular}}} &
          \multicolumn{2}{c|}{\cellcolor[HTML]{EFEFEF}\textbf{\begin{tabular}[c]{@{}c@{}}Smell \\ reduction rate\end{tabular}}} \\ \cline{2-7} 
        \rowcolor[HTML]{EFEFEF} 
        \multirow{-2}{*}{\cellcolor[HTML]{EFEFEF}\textbf{RefAgent}} &
          \multicolumn{1}{c|}{\cellcolor[HTML]{EFEFEF}{\color[HTML]{000000} \textbf{Median}}} &
          {\color[HTML]{000000} \textbf{Avg}} &
          \multicolumn{1}{c|}{\cellcolor[HTML]{EFEFEF}{\color[HTML]{000000} \textbf{Median}}} &
          {\color[HTML]{000000} \textbf{Avg}} &
          \multicolumn{1}{c|}{\cellcolor[HTML]{EFEFEF}{\color[HTML]{000000} \textbf{Median}}} &
          {\color[HTML]{000000} \textbf{Avg}} \\ \hline
        \textbf{\begin{tabular}[c]{@{}c@{}} GPT\end{tabular}} &
          \multicolumn{1}{c|}{90} &
          86.8 &
          \multicolumn{1}{c|}{87} &
          89.6 &
          \multicolumn{1}{c|}{52.5} &
          53.75 \\ \hline
        \textbf{\begin{tabular}[c]{@{}c@{}} Starcoder\end{tabular}} &
          \multicolumn{1}{c|}{85} &
          83.8 &
          \multicolumn{1}{c|}{84} &
          83.6 &
          \multicolumn{1}{c|}{50} &
          50.3 \\ \hline
        \textbf{\begin{tabular}[c]{@{}c@{}} DeepSeek-coder\end{tabular}} &
          \multicolumn{1}{c|}{90} &
          98.5 &
          \multicolumn{1}{c|}{88} &
          91.8 &
          \multicolumn{1}{c|}{53.5} &
          52.54 \\ \hline
        \rowcolor[HTML]{EFEFEF} 
        \textbf{p-value} &
          \multicolumn{2}{c|}{\cellcolor[HTML]{EFEFEF}0.053} &
          \multicolumn{2}{c|}{\cellcolor[HTML]{EFEFEF}0.064} &
          \multicolumn{2}{c|}{\cellcolor[HTML]{EFEFEF}0.252} \\ \hline
    \end{tabular}
    }
    \vspace{-2mm}
\end{table}

\subsubsection{\textbf{Findings}}

\textbf{RefAgent exhibits high unit test and compilation pass rates, indicating that a multi-agent approach can preserve functionality and syntactic correctness, while effectively reducing code smells.}

As shown in Table \ref{tab:refactoring_comparison}, RefAgent achieves a median unit test pass rate of 90\% and 90.5\% when using GPT-4o and DeepSeekCoder, respectively, while RefAgent-StarCoder achieves around 85\%. Compilation pass rates follow a similar trend, with RefAgent-DeepSeeker-coder and RefAgent-GPT attaining a median of 87\% and 88\%, respectively, compared to 84\% for RefAgent-StarCoder. The consistently high unit test pass rates and compilation success rates suggest that our multi-agent approach effectively preserves functionality across refactored projects, ensuring that structural improvements do not compromise correctness and syntactic integrity, while achieving an average code smell reduction of 52.5\%, 53.5\%, and 50\% for RefAgent using GPT-4o, DeepSeek-Coder, and StarCoder, respectively.

\begin{figure}[H]
    \centering
    \vspace{-2mm}
    \includegraphics[width=0.9\linewidth]{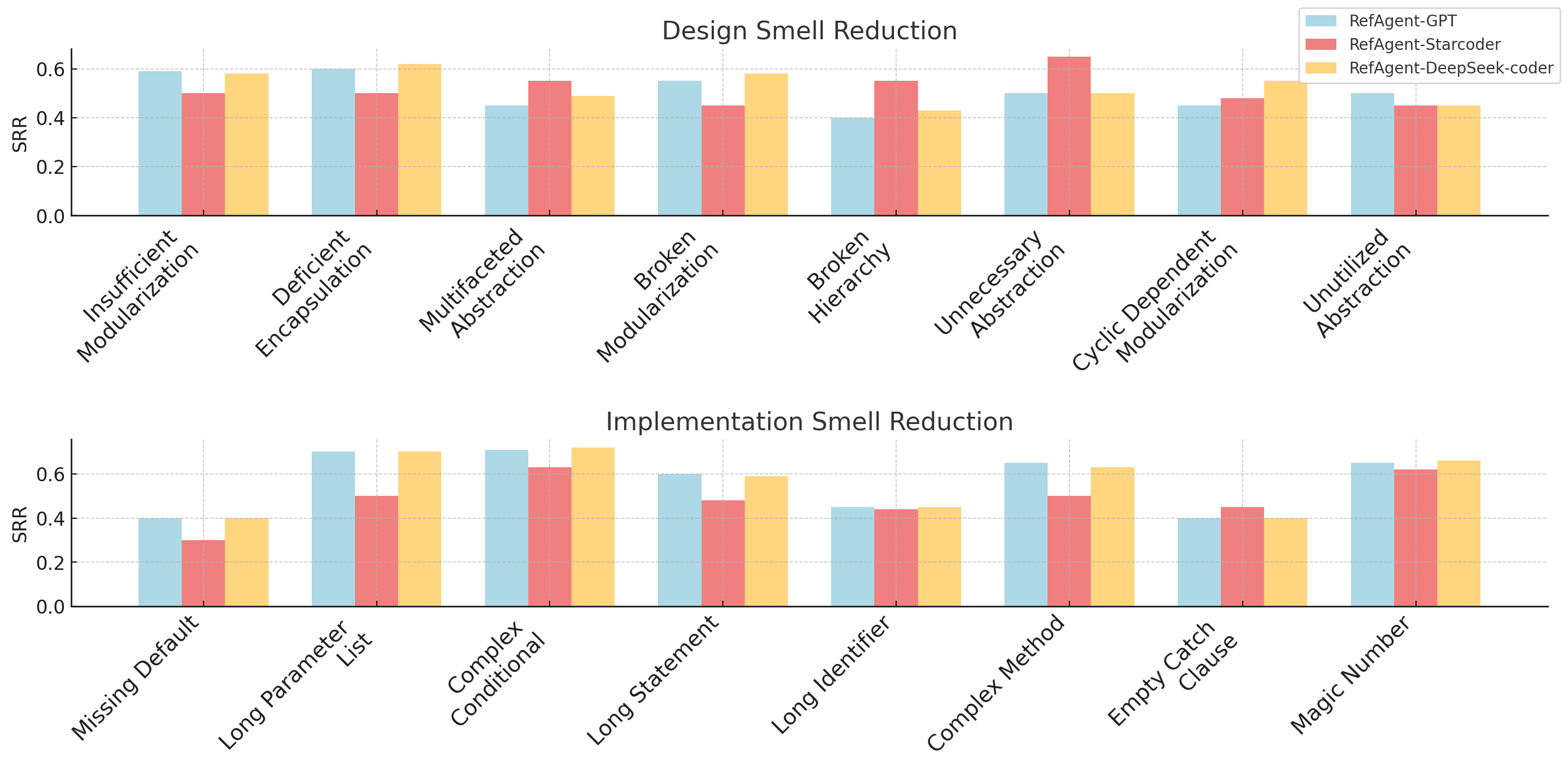}
    \vspace{-2mm}
    \caption{Comparison of Design and Implementation Code Smell Reduction Rate (SRR) for RefAgent-GPT, RefAgent-DeepSeek-coder and RefAgent-Starcoder.}
    \label{fig:code_smell_type}
    \vspace{-2mm}
\end{figure}

\textbf{RefAgent operates agnostically
to the underlying LLM. When evaluating unit test pass rates, compilation success, and smell reduction rates, no statistically significant differences were observed between GPT-4o, DeepSeek-Coder, and StarCoder (p > 0.05).}
Figure \ref{fig:code_smell_type} shows the effectiveness of RefAgent using GPT-4o, DeepSeek-Coder, and StarCoder in reducing different categories of code smells.  RefAgent using GPT-4o and DeepSeek-coder are more effective in improving modularization and reducing complexity, which may suggest that they have a better structural understanding. In contrast, RefAgent-Starcoder is better at handling abstraction and implementing simpler fixes at the implementation level (e.g., removing unnecessary elements, such as magic numbers).



\begin{figure}[H]
    \vspace{-2mm}
    \centering    \includegraphics[width=0.98\linewidth, ]{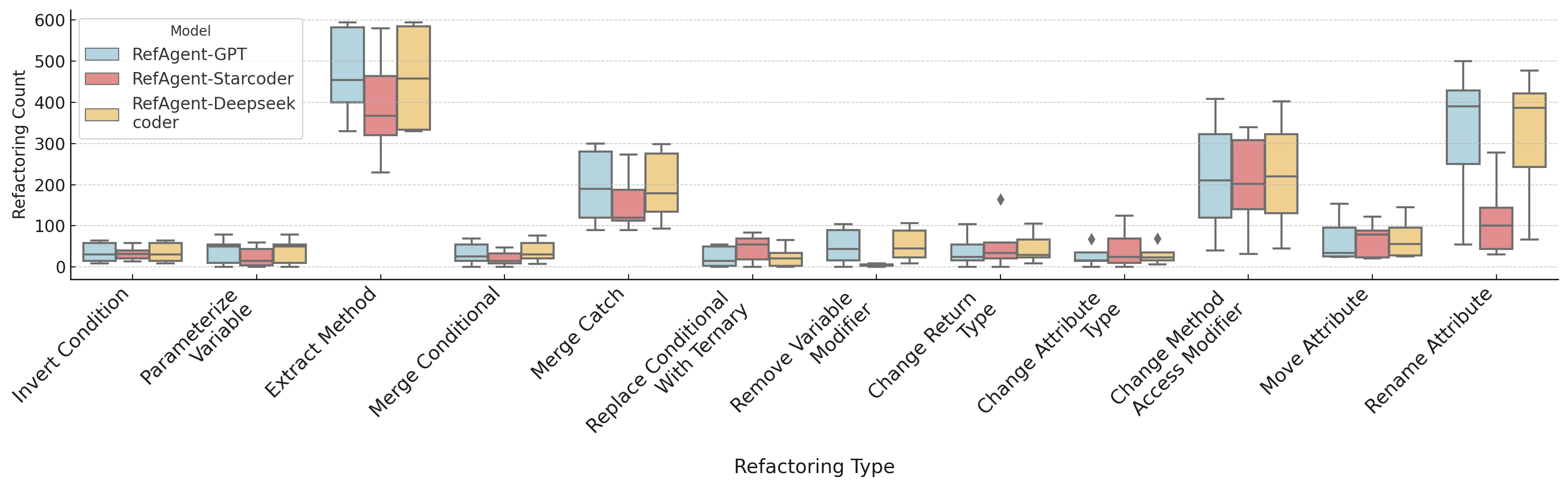}
    \vspace{-4mm}
    \caption{Comparison of Top 12 Refactoring Types counts across 8 projects}
    \label{fig:refactoring}
    \vspace{-3mm}
\end{figure}

Next, we use RefactoringMiner to identify the refactoring types applied by RefAgent. A total of 23 refactoring types were detected across projects. Figure \ref{fig:refactoring} shows the distribution of the top 12 refactoring types that RefAgent performs. We show that GPT-4o, DeepSeek-Coder, and Starcoder have comparable refactoring behaviors in Extract Method, Invert Condition, Parameterize Variable, and Merge Conditional. However, RefAgent using GPT-4o and DeepSeek-Coder applies Rename Attribute and Change Method Access Modifier 30\% more than Starcoder. In contrast, RefAgent-Starcoder takes a more conservative stance, performing Remove Variable Modifier and Replace Conditional With Ternary 60\% less than RefAgent-GPT or RefAgent-DeepSeek-Coder, suggesting a reluctance to modify variables and conditional structures.

RefAgent’s refactoring behavior varies depending on the underlying LLM, with GPT-4o and DeepSeek-Coder favoring more structural changes while Starcoder adopts a more conservative approach, suggesting that the choice of LLM can influence the balance between code transformation and preservation in 
automated refactoring. 

\begin{tcolorbox}
\vspace{-2mm}
RefAgent exhibits a high median unit test rate (median 90\%) and compilation pass rates (median 87\%), indicating that a multi-agent approach can preserve functionality and syntactic correctness while effectively reducing code smells. RefAgent operates agnostically to the underlying LLM. Moreover, RefAgent performs 23 types of refactoring.
\vspace{-4mm}
\end{tcolorbox}

\subsection{RQ2: How effective is our approach in identifying refactoring opportunities and improving software quality compared to search-based techniques and developers?}

\subsubsection{\textbf{Motivation}}

RefAgent relies on the reasoning capabilities of LLM-based agents, which can reason over the given context, generate, and validate refactorings through dynamic workflows. Specifically, the Context-aware Planner Agent is responsible for identifying refactoring opportunities and assigning the proper refactorings for the various code regions (e.g., method, field, variable) of the target class. To examine the effectiveness of RefAgent, we compare its ability to identify and apply meaningful refactorings against both automated baselines and developer-applied changes.

\subsubsection{\textbf{Approach}}
To evaluate RefAgent's ability in identifying refactoring opportunities, we compare its refactoring selections with those made by developers and RefGen~\cite{MORALES201825}, a search-based refactoring algorithm that was shown to effectively identify refactoring opportunities. The objective is to determine whether RefAgent applies refactorings in the same class, method, and location as developers and RefGen.

First, for every class refactored by RefAgent, we extract refactoring locations and types from our 8 projects using Refactoring Miner as discussed in Section~\ref{sec:refactoring}. Second, we execute the search-based refactoring tool RefGen on the projects, from which we extract the refactored classes. Finally, in the case for developers, we select the next release for each of our eight projects, then run RefactoringMiner on it to collect historical refactoring commits that contain the refactored classes, methods as well as the types. 



We present two scenarios:

\noindent \textbf{Scenario 1}: We determine a match between RefAgent's and developers' refactorings by comparing the class name, method name, line of code range, and refactoring type: if all attributes align, it is a match; otherwise, it is not a match. 

\noindent \textbf{Scenario 2}: Since RefGen tool does not provide location and code change details, we determine a match if RefGen and RefAgent apply the same refactoring type to the same class and method. 



Finally, we compute Precision, Recall, and F1-score, as discussed in Section~\ref{sec:refactoring}, to quantify the degree of alignment, determining whether RefAgent more closely follows developer decisions or search-based approaches.

\subsubsection{\textbf{Findings}}


\textbf{RefAgent effectively identifies refactoring opportunities, achieving strong alignment with both developers (78\% median across our subject systems) and RefGen (72\% median across our subject systems)}.

\begin{table}[H]
\label{label:precision_recall}
\vspace{-2mm}
\caption{Median values of Precision, Recall, F1-score of RefAgent compared to developers vs Refgen}
\label{tab:precision_recall}

\vspace{-3mm}
\scalebox{0.69}{
\begin{tabular}{|lcccccc|}
\hline
\rowcolor[HTML]{EFEFEF} 
\multicolumn{1}{|l|}{\cellcolor[HTML]{EFEFEF}} &
  \multicolumn{3}{c|}{\cellcolor[HTML]{EFEFEF}Developer} &
  \multicolumn{3}{c|}{\cellcolor[HTML]{EFEFEF}RefGen} \\ \cline{2-7} 
\rowcolor[HTML]{EFEFEF} 
\multicolumn{1}{|l|}{\multirow{-2}{*}{\cellcolor[HTML]{EFEFEF}Refagent}} &
  \multicolumn{1}{l|}{\cellcolor[HTML]{EFEFEF}\# Precision} &
  \multicolumn{1}{l|}{\cellcolor[HTML]{EFEFEF}\# Recall} &
  \multicolumn{1}{l|}{\cellcolor[HTML]{EFEFEF}\# F1-Score} &
  \multicolumn{1}{l|}{\cellcolor[HTML]{EFEFEF}\# Precision} &
  \multicolumn{1}{l|}{\cellcolor[HTML]{EFEFEF}\# Recall} &
  \multicolumn{1}{l|}{\cellcolor[HTML]{EFEFEF}\# F1-Score} \\ \hline
\multicolumn{1}{|l|}{\cellcolor[HTML]{EFEFEF}GPT-4o} &
  \multicolumn{1}{c|}{78} &
  \multicolumn{1}{c|}{81} &
  \multicolumn{1}{c|}{80} &
  \multicolumn{1}{c|}{75} &
  \multicolumn{1}{c|}{70} &
  72 \\ \hline
\multicolumn{1}{|l|}{\cellcolor[HTML]{EFEFEF}DeepSeek-coder} &
  \multicolumn{1}{c|}{79} &
  \multicolumn{1}{c|}{81} &
  \multicolumn{1}{c|}{80} &
  \multicolumn{1}{c|}{75} &
  \multicolumn{1}{c|}{70} &
  72 \\ \hline
\multicolumn{1}{|l|}{\cellcolor[HTML]{EFEFEF}StarCoder} &
  \multicolumn{1}{c|}{72} &
  \multicolumn{1}{c|}{73} &
  \multicolumn{1}{c|}{73} &
  \multicolumn{1}{c|}{69} &
  \multicolumn{1}{c|}{68} &
  69 \\ \hline
\rowcolor[HTML]{EFEFEF} 
\multicolumn{1}{|l|}{\cellcolor[HTML]{EFEFEF}\textbf{Median}} &
  \multicolumn{1}{c|}{\cellcolor[HTML]{EFEFEF}\textbf{78}} &
  \multicolumn{1}{c|}{\cellcolor[HTML]{EFEFEF}\textbf{81}} &
  \multicolumn{1}{c|}{\cellcolor[HTML]{EFEFEF}\textbf{80}} &
  \multicolumn{1}{c|}{\cellcolor[HTML]{EFEFEF}\textbf{75}} &
  \multicolumn{1}{c|}{\cellcolor[HTML]{EFEFEF}\textbf{70}} &
  \textbf{72} \\ \hline
\multicolumn{7}{|l|}{\# Median value, rounded number} \\ \hline
\end{tabular}
}
\vspace{-2mm}
\end{table}

In table~\ref{tab:precision_recall}, we compare the precision, recall, and F1-score of RefAgent against developers and RefGen. The results highlight that RefAgent achieves a high median recall of 81\% and a median F1-score of 80\% when compared to developers, demonstrating its ability to identify refactoring opportunities and mimicking developers' intuition. Precision remains slightly lower with a median value of 78\%, suggesting that while RefAgent identifies a broad range of changes, some improvement may still be needed. Additionally, we observe that RefAgent shows strong alignment with the optimized search-based tool RefGen, achieving a median F1-score of 72\%. When using GPT-4o, DeepSeek-Coder, and StarCoder, RefAgent maintains a consistent performance with F1-score 80\%, 80\% and 69\% respectively, demonstrating its ability to generalize well across different LLM architectures. Since GPT-4o, being a closed-source model, faces the risk of data leakage. We therefore rely on StarCoder2 to determine the strength of our approach in a more controlled setting.

While precision, recall, and F1-score provide useful insights into RefAgent’s effectiveness, their interpretation should be approached with nuance. These metrics assume that developers and RefGen serve as definitive references, yet they may also overlook meaningful refactorings that RefAgent successfully identifies. As a result, some misalignment does not necessarily indicate that RefAgent produced incorrect or suboptimal results, particularly if the applied refactorings lead to improved code quality. Given that search-based approaches aim to generate optimal solutions, it is important to acknowledge that multiple valid refactoring solutions may exist. Therefore, rather than viewing misalignments strictly as errors, they should be carefully analyzed to assess whether RefAgent is capturing valuable transformations beyond those identified by developers and RefGen. In future work, we will further investigate this aspect to better understand the impact of RefAgent’s refactorings.

\begin{table}[H]
\vspace{-3mm}
\centering
\caption{QMOOD Improvement Rates (IR) values across different attributes for RefAgent vs RefGen.}
\vspace{-2mm}
\scalebox{0.77}{
\begin{tabular}{|l|r|r|r|r|}
\hline
\rowcolor[HTML]{EFEFEF} 
\textbf{QMOOD} &
  \makecell{\textbf{RefAgent}\\\textbf{GPT-4o}} &
  \makecell{\textbf{RefAgent}\\\textbf{Starcoder}} &
  \makecell{\textbf{RefAgent}\\\textbf{Deepseek-coder}} &
  \textbf{RefGen} \\ \hline
\textbf{Reusability}       & 8.16                  & 5.48                        & 8.19                              & 16.5            \\ \hline
\textbf{Flexibility}       & -0.62                 & -0.35                       & -0.63                             & -5.9            \\ \hline
\textbf{Understandability} & -14.34                & -13.47                      & -14.44                            & -19.7           \\ \hline
\textbf{Functionality}     & 4.75                  & 0.49                        & 4.75                              & 0               \\ \hline
\textbf{Extendibility}     & 0                     & 0                           & 0                                 & 5.9             \\ \hline
\textbf{Effectiveness}     & 4.28                  & 3.39                        & 4.29                              & -9.7           \\ \hline
\textbf{p-value}           & 0.843                 & 0.687                       & 0.852                                 & -              \\ \hline
\end{tabular}
}
\vspace{-3mm}
\label{tab:qmmod_search_based}
\end{table}

From Table~\ref{tab:qmmod_search_based}, we observe that RefAgent improves reusability, understandability, and functionality compared to RefGen. Given the metric definitions in Table~\ref{tab:qmood}, these improvements indicate that RefAgent enhances method cohesion, which aligns with our findings in RQ1~\ref{RQ1}, where RefAgent frequently applies Extract Method refactorings or modifies method access modifiers.

Additionally, while understandability is lower in RefAgent than in RefGen, the difference remains comparable. This can be attributed to RefAgent’s effectiveness in reducing complexity compared to RefGen, as lower complexity can lead to reduced understandability despite improved cohesion.

Moreover, compared to RefGen, RefAgent shows an increase in effectiveness and a slight increase in flexibility. This can be explained by RefAgent’s improvement in the number of polymorphic methods, as polymorphism positively impacts both metrics by enhancing method adaptability and reusability within the design.

Overall, RefAgent’s structured refinement process through multi-agent collaboration demonstrates refactoring improvements that are competitive and comparable (p-value $\geq$ 0.05) with the search-based optimization approach of RefGen.

\begin{tcolorbox}
\vspace{-2mm}
RefAgent presents a fully autonomous means to identify refactoring opportunities and improve software quality, with performance comparable to developers (median F1-score 80\%) and RefAgent (median F1-score 72\%). Moreover, RefAgent shows comparable performance in QMOOD improvements compared to RefGen (p-value>0.05)

\vspace{-2mm}
\end{tcolorbox}
\subsection{RQ3: What is the contribution of each component of our framework?}

\subsubsection{\textbf{Motivation}}
While multi-agent architectures offer modularity and the ability to decompose complex tasks, they also introduce design and orchestration overhead. Without a clear understanding of the role of each component, it becomes difficult to justify this complexity or optimize the framework. Prior work in software engineering and LLM-based systems highlights the importance of component-level analysis and ablation studies to quantify the effectiveness of individual elements~\cite{nunez2024autosafecodermultiagentframeworksecuring}.

\subsubsection{\textbf{Comparison of RefAgent with single LLM-based approaches.}}


To assess the performance of RefAgent, we compare its refactored outputs against single-agent approaches, which are based on previous work \cite{cordeiro2024empiricalstudycoderefactoring}, in terms of software quality improvements. The comparison focuses on key metrics used in RQ1, RQ2 ( unit test Pass Rate, compilation success rate, Smell Reduction Rate (SRR), and QMOOD metrics). For the single-agent baseline, we evaluate the effectiveness of an LLM-based refactoring agent given the same context as RefAgent to ensure a fair comparison. Similarly to previous work, we use for the unit test pass rate evaluation the pass @k metric defined as follows:

\begin{itemize}
    \item Pass@1: We generate a single refactored solution for the target class. If it passes the unit test, it is considered a successful refactoring under the Pass@1 criterion.
    \item Pass@3: We generate three refactored solutions for each target class. Each solution is independently evaluated by running the corresponding unit tests. If at least one of these three refactored versions passes all unit tests, the refactoring is considered successful under pass@3.
\end{itemize}

This setup ensures that RefAgent is fairly compared fairly to a single-agent LLM. Similarly to RQ1 and RQ2, we evaluate this RQ on all of the classes of our 8 subject systems presented in Table \ref{tab:projects}.

\subsubsection{\textbf{Findings}}

\textbf{RefAgent significantly outperforms the single agent approach in terms of unit test pass rate, compilation pass rates, and code quality improvement across 8 software projects.}

From Table ~\ref{tab:single_agent}, we can identify that incorporating feedback loops in a RefAgent and decoupling the planner and execution agents enhances the robustness and correctness of code refactoring, leading to better overall software quality. In fact, RefAgent shows a high improvement rate across all metrics with GPT4o, Deepseek-Coder and Starcoder2, when compared to single-agent Pass@1 and Pass@3, demonstrating that the approach is not sensitive to the choice of LLM. For instance, in GPT-4o and DeepSeek-Coder, RefAgent achieves a unit test pass rate of 90\% compared to 44.5\% in Pass@1, while in Starcoder2, RefAgent improves the compilation pass rate to 84\%, significantly outperforming single-LLM agents in the Pass@1 scenario. These improvements, backed by statistically significant p-values 
confirm the effectiveness of RefAgent when compared to single-agent approaches.
\begin{table}[H]
\caption{Comparison of test pass rates and compilation pass rates for single agent approach vs RefAgent.}
\vspace{-2mm}
\scalebox{0.75}{

\begin{tabular}{|cc|cc|cc|cc|}
\hline
\rowcolor[HTML]{EFEFEF} 
\multicolumn{2}{|c|}{\cellcolor[HTML]{EFEFEF}} &
  \multicolumn{2}{c|}{\cellcolor[HTML]{EFEFEF}\textbf{\begin{tabular}[c]{@{}c@{}}Unit test\\ pass rate\end{tabular}}} &
  \multicolumn{2}{c|}{\cellcolor[HTML]{EFEFEF}\textbf{\begin{tabular}[c]{@{}c@{}}Compilation \\ pass rate\end{tabular}}} &
  \multicolumn{2}{c|}{\cellcolor[HTML]{EFEFEF}\textbf{\begin{tabular}[c]{@{}c@{}}Smell \\ reduction rate\end{tabular}}} \\ \cline{3-8} 
\rowcolor[HTML]{EFEFEF} 
\multicolumn{2}{|c|}{\multirow{-2}{*}{\cellcolor[HTML]{EFEFEF}\textbf{Refactoring}}} &
  \multicolumn{1}{c|}{\cellcolor[HTML]{EFEFEF}\textbf{Median}} &
  \textbf{Avg} &
  \multicolumn{1}{c|}{\cellcolor[HTML]{EFEFEF}\textbf{Median}} &
  \textbf{Avg} &
  \multicolumn{1}{c|}{\cellcolor[HTML]{EFEFEF}\textbf{Median}} &
  \textbf{Avg} \\ \hline

\multicolumn{1}{|c|}{} & \textbf{RefAgent} & \multicolumn{1}{c|}{90} & 86.8 & \multicolumn{1}{c|}{87} & 89.6 & \multicolumn{1}{c|}{52.5} & 53.75 \\ \cline{2-8} 
\multicolumn{1}{|c|}{} & \textbf{Pass@1} & \multicolumn{1}{c|}{44.5} & 44.3 & \multicolumn{1}{c|}{48} & 48.4 & \multicolumn{1}{c|}{38.1} & 41.2 \\ \cline{2-8} 
\multicolumn{1}{|c|}{\multirow{-3}{*}{\textbf{GPT}}} & \textbf{Pass@3} & \multicolumn{1}{c|}{56} & 60 & \multicolumn{1}{c|}{62} & 64.6 & \multicolumn{1}{c|}{42.5} & 43.1 \\ \hline
\multicolumn{1}{|c|}{\cellcolor[HTML]{EFEFEF}\textbf{P-value}} & \textbf{-} & \multicolumn{2}{c|}{\cellcolor[HTML]{EFEFEF}0.001} & \multicolumn{2}{c|}{\cellcolor[HTML]{EFEFEF}0.007} & \multicolumn{2}{c|}{\cellcolor[HTML]{EFEFEF}0.039} \\ \hline

\multicolumn{1}{|c|}{} & \textbf{RefAgent} & \multicolumn{1}{c|}{85} & 83.8 & \multicolumn{1}{c|}{84} & 83.6 & \multicolumn{1}{c|}{50} & 50.3 \\ \cline{2-8} 
\multicolumn{1}{|c|}{} & \textbf{Pass@1} & \multicolumn{1}{c|}{33} & 35.2 & \multicolumn{1}{c|}{45} & 45.4 & \multicolumn{1}{c|}{37.5} & 39 \\ \cline{2-8} 
\multicolumn{1}{|c|}{\multirow{-3}{*}{\textbf{Starcoder}}} & \textbf{Pass@3} & \multicolumn{1}{c|}{52.5} & 51 & \multicolumn{1}{c|}{56} & 59 & \multicolumn{1}{c|}{48.3} & 50 \\ \hline
\multicolumn{1}{|c|}{\cellcolor[HTML]{EFEFEF}\textbf{P-value}} & \textbf{-} & \multicolumn{2}{c|}{\cellcolor[HTML]{EFEFEF}0.02} & \multicolumn{2}{c|}{\cellcolor[HTML]{EFEFEF}0.007} & \multicolumn{2}{c|}{\cellcolor[HTML]{EFEFEF}0.039} \\ \hline

\multicolumn{1}{|c|}{} & \textbf{RefAgent} & \multicolumn{1}{c|}{90.5} & 87.3 & \multicolumn{1}{c|}{88.5} & 89.2 & \multicolumn{1}{c|}{53.5} & 54.1 \\ \cline{2-8}
\multicolumn{1}{|c|}{} & \textbf{Pass@1} & \multicolumn{1}{c|}{44.5} & 45.2 & \multicolumn{1}{c|}{50} & 49.5 & \multicolumn{1}{c|}{40.5} & 42.3 \\ \cline{2-8}
\multicolumn{1}{|c|}{\multirow{-3}{*}{\makecell{\textbf{DeepSeek}\\\textbf{coder}}}} & \textbf{Pass@3} & \multicolumn{1}{c|}{58} & 61 & \multicolumn{1}{c|}{63} & 65.2 & \multicolumn{1}{c|}{49.2} & 51.6 \\ \hline
\multicolumn{1}{|c|}{\cellcolor[HTML]{EFEFEF}\textbf{P-value}} & \textbf{-} & \multicolumn{2}{c|}{\cellcolor[HTML]{EFEFEF}0.004} & \multicolumn{2}{c|}{\cellcolor[HTML]{EFEFEF}0.005} & \multicolumn{2}{c|}{\cellcolor[HTML]{EFEFEF}0.039} \\ \hline
\end{tabular}
}
\vspace{-3mm}
\label{tab:single_agent}
\end{table}

\begin{figure}[H]
    \vspace{-2mm}
    \centering
    \includegraphics[width=\linewidth]{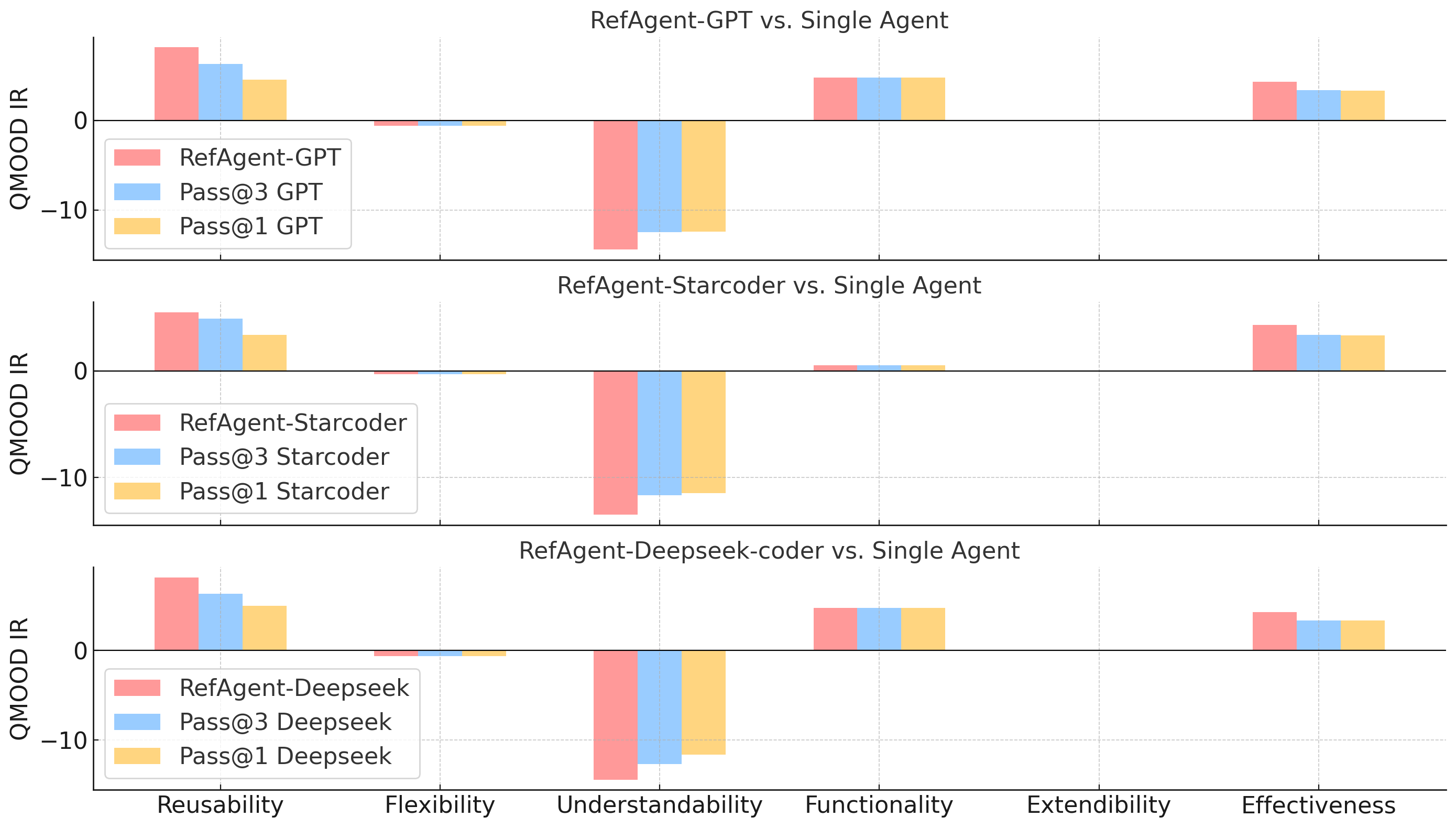}
    \caption{Improvement rate of QMood Metrics: RefAgent-GPT vs single agents and RefAgent-Starcoder vs Single agents. The Wilcoxon t-test yields a p-value of 0.046}
    \vspace{-3mm}
    \label{fig:single_agent_qmood}
    \vspace{-1mm}
\end{figure}

\textbf{RefAgent consistently and significantly with (p-value< 0.05) outperforms the single-agent approaches across all QMOOD metrics, demonstrating that the multi-agent approach is important to improve software quality.} From figure~\ref{fig:single_agent_qmood}, we show that RefAgent with subject LLMs shows similar improvements over the single-agent counterpart. This suggests that the effectiveness of RefAgent is not dependent on the specific LLM used but rather on the multi-agent collaboration itself. This indicates that RefAgent enhances reusability and functionality attributes, and overall code quality beyond what a single-agent model can achieve alone. 
The Figure~\ref{fig:single_agent_qmood} further suggests that multi-agent coordination introduces something more to the refactoring process, leading to better software design decisions.

\subsubsection{\textbf{Ablation study}}
We conduct an ablation study by systematically removing key components of RefAgent and analyzing its performance. Specifically, we assess the impact of eliminating context retrieval, dependency analysis, and the code search module from the framework. We analyze the unit test pass rates and compilation pass rates across each scenario.

Furthermore, we analyze the impact of the feedback loop by evaluating the unit test pass rate and compilation pass rate over 20 iterations. This will help determine how the iterative refinement process influences model performance and stability over 20 iterations.

By conducting these experiments, we aim to isolate the contribution of each component and understand their role in enhancing the quality and effectiveness of the generated refactoring.

\subsubsection{\textbf{Findings}}
\begin{figure}[H]
\vspace{-4mm}
    \centering    \includegraphics[width=0.9\linewidth]{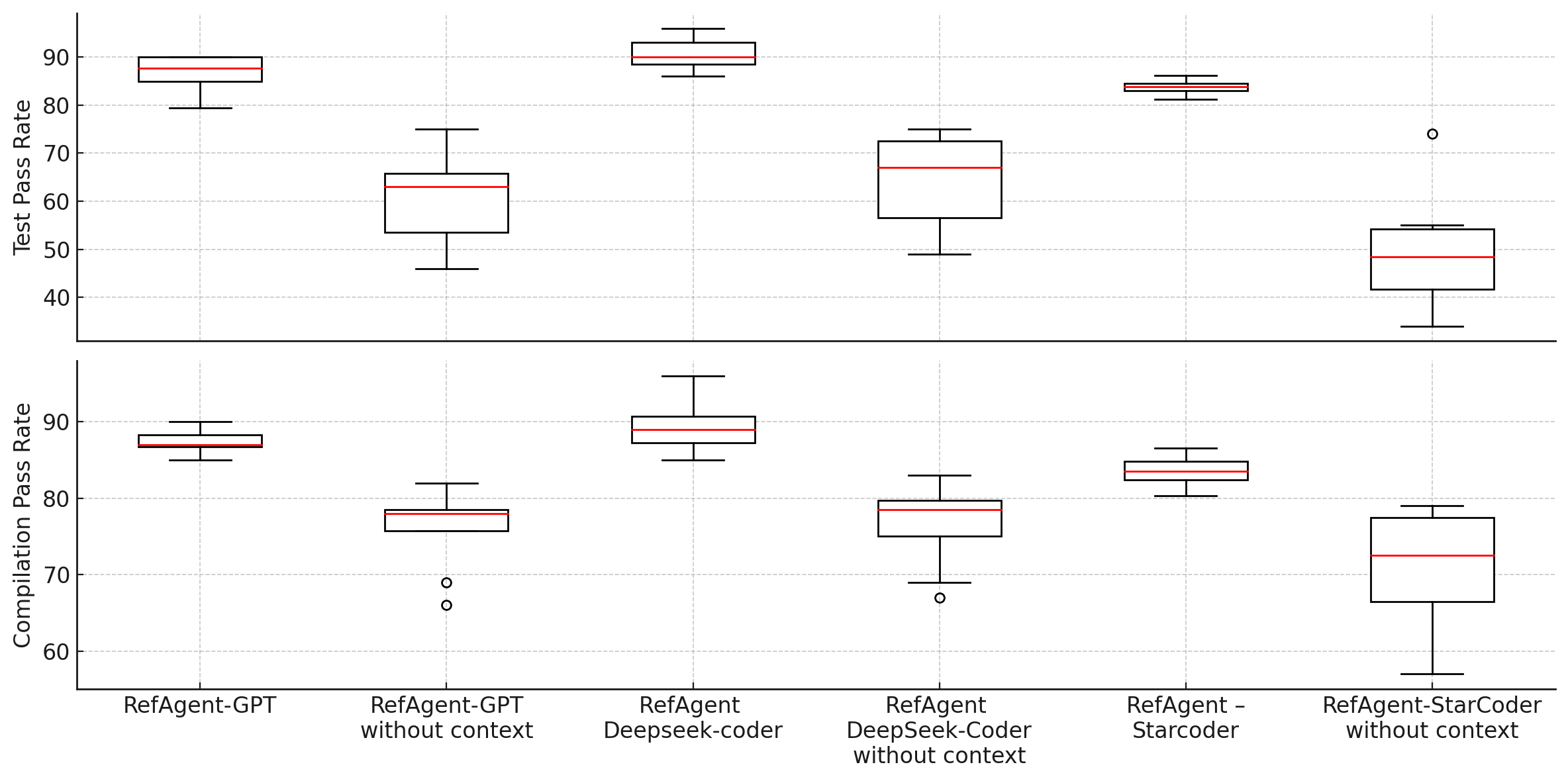}
    \vspace{-2mm}
    \caption{Distribution of test pass rates and compilation pass rates for RefAgent with and without context, highlighting variations in performance across different configurations.}
    \vspace{-3mm}
    \label{fig:ablation-metrics}
   
\end{figure}

\begin{figure}[H]

    \centering
    \includegraphics[width=0.9\linewidth]{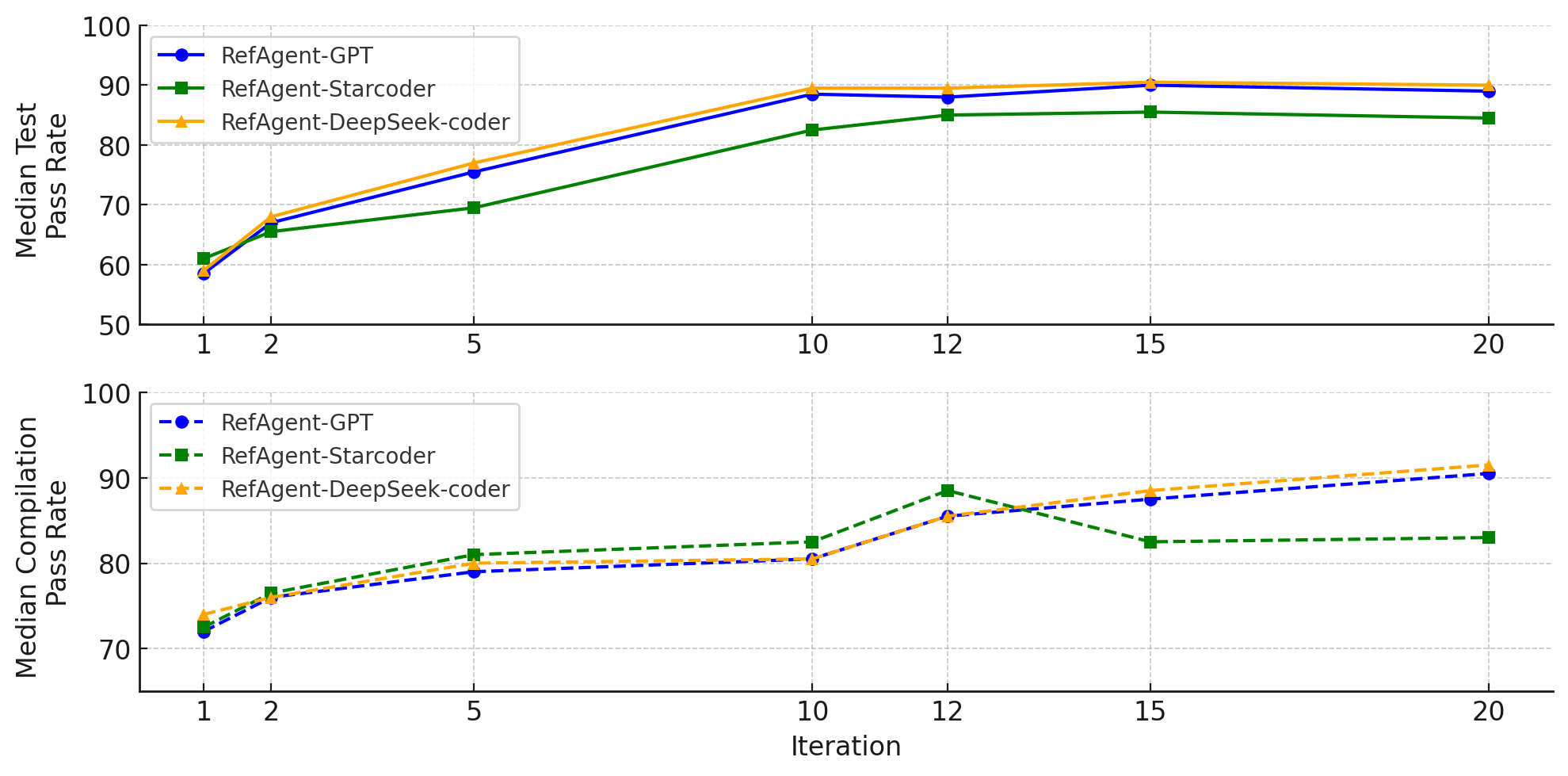}
    \vspace{-3mm}
     \caption{Test and Compilation Pass Rate Across Iterations (GPT-4o vs. Starcoder vs. DeepSeek-Coder)}
    \label{fig:test_rate_itter}
    \vspace{-2mm}
\end{figure}

\textbf{Context retrieval is essential for maintaining high test and compilation pass rates in RefAgent, as its removal leads to significant performance degradation and increased variability.} Figure~\ref{fig:ablation-metrics} presents results of the ablation study on context retrieval components, comparing RefAgent-GPT, RefAgent-DeepSeek-Coder, and RefAgent-Starcoder with and without context. Figure~\ref{fig:ablation-metrics} shows that removing context leads to a significant drop in both unit test pass rates and compilation pass rates, highlighting the critical role of context in ensuring refactoring effectiveness. RefAgent using subject LLMs exhibits high performance when context is included, but without it, performance becomes highly unstable, with increased variance and a lower median pass rate.

\textbf{RefAgent progressively enhances the unit test pass rates across 20 iterations with the 3 subject LLMs.} From Figure~\ref{fig:test_rate_itter}, one can observe that the median test and compilation pass rate steadily increase with each iteration, stabilizing at 90\% and 89\% after 12 iterations, demonstrating the compounding benefits of iterative refinement. This iterative approach allows RefAgent to identify and correct errors in refactoring recommendations and improve code structure. Ultimately, iterating is essential for achieving stable and high-quality software outputs for RefAgent.

\begin{tcolorbox}
\vspace{-2mm}
RefAgent significantly outperforms the single-agent approach in terms of unit test pass rate, compilation pass rates, and QMOOD quality improvements across 8 software projects. This demonstrates the usefulness of a multi-agent approach over a single agent approach when improving software quality. Furthermore, the ablation study highlights the importance of context retrieval and iterative refinement for maintaining high test and compilation pass rates with agentic software refactoring. 
\vspace{-2mm}
\end{tcolorbox}  

\section{Discussion}\label{sec:discussion}

In this section, we ensure that our evaluation is not adversely affected by tool-specific limitations. Thus, we perform manual validation of a representative sample of the outputs from DesigniteJava, EvoSuite, and RefactoringMiner.

\begin{figure}[H]
\vspace{-3mm}
    \centering    \includegraphics[width=\linewidth]{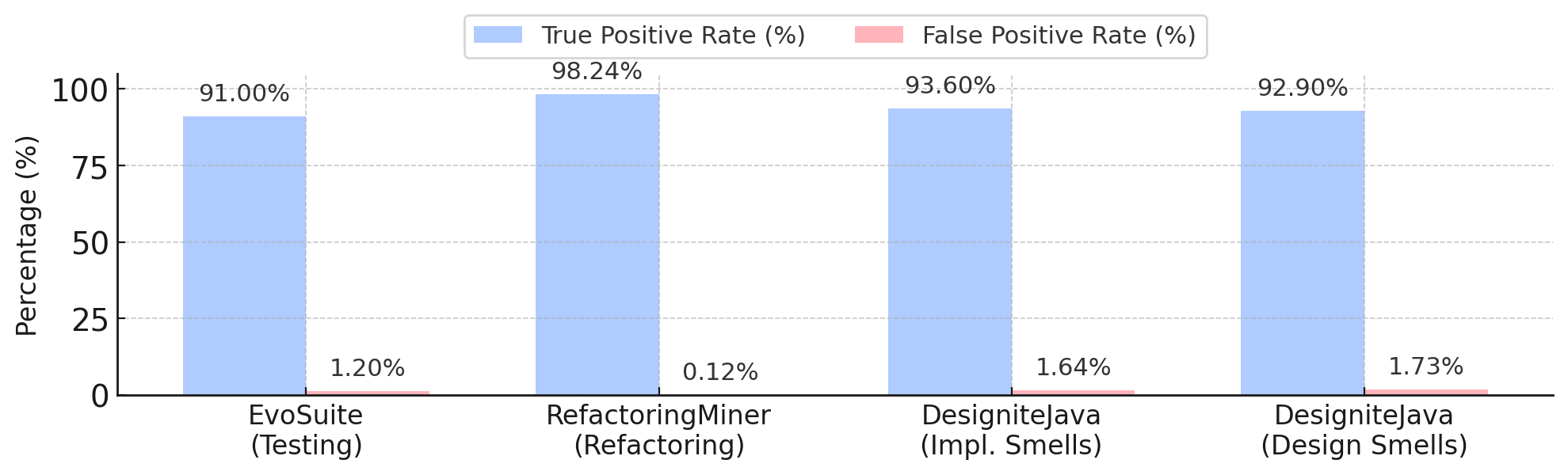}
    \vspace{-5mm}
    \caption{True Positive and False Positive rates of the tools used in our study, based on manual validation of a statistically representative sample.}
    \vspace{-2mm}
    \label{fig:manual_analysis}
    \vspace{-2mm}
\end{figure}




\noindent \textit{\textbf{Manual analysis of the Evosuite's generated tests.}}
\noindent In RefAgent, EvoSuite is invoked by the Tester agent to generate tests for the refactored classes when developer-written tests are missing or insufficient.

To assess the reliability of Evosuite, we randomly select a statistically significant sample, averaging 374 classes per project from each of the eight open-source projects (95\% confidence, 5\% margin of error). For each class, we manually review all generated test cases to determine whether they were True Positives (TP), i.e., semantically correct and passing, or False Positives (FP), i.e., incorrect but still passing. As shown in Figure~\ref{fig:manual_analysis}, an average of 91\% of tests were TP, closely aligning with EvoSuite’s reported accuracy and prior studies~\cite{cordeiro2024empiricalstudycoderefactoring, shamshiri2015evosuite, 10.1145/3526072.3527526}. The average FP rate was only 1.2\%, confirming the low risk of misleading results. These findings support the reliability of EvoSuite within RefAgent. While FPs are a known limitation in automated test generation, their minimal presence suggests a limited impact on decision-making. Future work may incorporate mutation testing or enhanced test oracles to better validate behavioral correctness. Additionally, leveraging LLM-as-a-judge approaches can help assess semantic soundness beyond pass/fail outcomes~\cite{gu2025surveyllmasajudge}.

\noindent \textit{\textbf{Impact of False Positives in RefAgent Evaluation.}}
\noindent While static tools such as DesigniteJavaonly detect code smells based on heuristic rules or pattern matching, RefAgent orchestrates an end-to-end workflow that includes planning, transformation, validation, and iterative correction through reasoning and collaboration, which traditional static tools are not equipped to perform. We use tools like DesigniteJava and RefactoringMiner for evaluation purposes. However, the planner agent does use the tool DesigniteJava only to extract low-level code metrics to augment the prompt and enrich context, which can be replaced with any available tool that computes code metrics, yet the core decision-making in the planner agent is driven by agent reasoning, enabling adaptability and generalization across diverse codebases and scenarios. This design choice prioritizes autonomy, interpretability, and the ability to evolve independently of static tool limitations.

\noindent To ensure the integrity of our evaluation, we manually analyzed a randomly-selected, statistically representative sample of \textbf{380 refactored classes} across eight projects, applying the same sample to both refactoring and code smell detection.

For refactoring detection, as shown in Figure~\ref{fig:manual_analysis}, RefactoringMiner yielded a \textbf{TP rate of 98.24\%} and a \textbf{FP rate of just 0.12\%}, confirming its reliability of the tool being aligned with the results reported by the authors of RefactoringMiner \cite{9136878} However, a few false positives, typically benign cases like formatting changes being misclassified as refactorings, may introduce minor noise.

For code smells, as shown in Figure~\ref{fig:manual_analysis}, DesigniteJava reported a \textbf{TP rate of 93.6\%} (implementation smells) and \textbf{92.9\%} (design smells), with corresponding FP rates of \textbf{1.64\%} and \textbf{1.73\%} respectively. These results align with prior works~\cite{Sharma2024}. While generally low, these false positives can slightly overestimate the number of smells detected or removed, particularly in borderline cases. 

These findings highlight a broader challenge in automated evaluations: the trade-off between scalability and semantic precision. 
Since RefAgent operates as a fully automated framework, developers are not required to run static analysis tools before or after refactoring. Instead, we employ these tools solely for empirical evaluation, to quantitatively assess the improvements in code quality (e.g., reduction of code smells, refactoring types). The developer is not involved in this validation loop. In practice, RefAgent autonomously performs refactoring, testing, and validation using agent reasoning and tool-calling when needed. Therefore, the manual effort required from developers is significantly reduced. However, this paves the way for future work to introduce a human-in-the-loop review mechanism~\cite{takerngsaksiri2025humanintheloopsoftwaredevelopmentagents} that interacts with agents, which can further ensure reliability. An example is an agent-human collaboration mechanism in which it can request input at specific conversation rounds based on the configuration. The default user proxy agent should enable customizable involvement, allowing users to define how often and under what conditions human input is requested, including the option to skip providing input. These strategies offer a promising path toward balancing automation with trustworthiness in evaluating intelligent software engineering systems.

\section{Threats to Validity}\label{sec:threats}
In this section, we discuss potential threats to the validity of our study and the measures we have taken to mitigate them.

\textbf{\textit{External Validity.}} In this study, we focused exclusively on datasets from Apache projects, which may limit the generalizability of our findings across different software development environments. However, this choice ensures consistency with related work \cite{cordeiro2024empiricalstudycoderefactoring}, enabling direct comparison and enhancing the validity and comparability of our results. Future research should explore more diverse datasets to further validate and broaden the applicability of our findings.

\textbf{\textit{Internal Validity.}}
Our study is based on the latest version of DeepSeeker-coder-33b, StarCoder2-15B-Instruct-v0.1 and GPT4o, as available at the time of analysis. While our results indicate that RefAgent performs well agnostically from the used LLM, , our evaluation approach is designed to be adaptable and can be applied to future versions of StarCoder2, GPT, and other LLMs from different vendors. However, differences in hardware setups can impact model performance, potentially leading to different results.

While LLMs may hallucinate invalid code, RefAgent mitigates this through a multi-agent design with reflection loops and tools like EvoSuite for validation. Though EvoSuite may introduce false positives, we manually validated samples and found minimal impact. Also, our study is designed to apply refactored code only if existing developer-written tests and auto-generated tests pass.



\textbf{\textit{Construct Validity.}} 
Our assessment of refactoring quality primarily relies on code smell reduction, improvements in code quality metrics, and unit test pass rates. However, these metrics may not comprehensively capture all aspects of code quality. This limitation could impact the robustness of our conclusions across a broader spectrum of quality indicators. We utilized Rminer3.0 and DesigniteJava in Java code analysis to collect our metrics. However, different tools may yield varying results.

\section{Conclusion}\label{sec:conclusion}

This paper introduces RefAgent, a fully automated, multi-agent refactoring framework that aims to enhance software quality across multiple software quality dimensions. Our approach leverages specialized LLM-based agents equipped with tool calling capabilities to dynamically retrieve context, interact with the compilation and
testing environment, and perform more complex context-aware refactoring tasks without additional manual oversight. Through our extensive evaluation of eight real-world open-source Java projects, RefAgent not only achieves significant reductions in code smells and improvements in quality attributes but also preserves functionality, as demonstrated by high unit test and compilation pass rates. Future work should look into balancing quality improvements with potential trade-offs in design flexibility and understandability. To the best of our knowledge, this is the first multi-agent approach specifically designed for software refactoring. 
\section{Data Availability}
Our data, and the scripts necessary to replicate our work, is available, under an open license, using
the following link: \url{https://github.com/anonymAgent/RefAgent}

\newpage
\bibliographystyle{ACM-Reference-Format}
\bibliography{sample-base}










\end{document}